\documentclass[graphicx,11pt]{revtex4-2}

\usepackage{amsmath}
\usepackage[british]{babel}
\usepackage{graphicx}
\usepackage{amssymb}
\usepackage{amsthm}
 \usepackage{tikz}
 \usetikzlibrary{tikzmark}
\usepackage{helvet}

\bibstyle{aipnum4-2}

\usepackage{amssymb}
\usepackage{amsthm}
\DeclareMathOperator{\sign}{sgn}

\usepackage{amsmath}
\usepackage{amssymb}

\newcommand{\mx}[1]{\ifmmode\mathbf{#1}\else\textbf{#1}\fi}

\newcommand{\labelfont}{\fontsize{9}{9}\selectfont}
\newcommand{\apanel}[1]{\textbf{\labelfont\textit{#1}}}

\newsavebox{\figbox}
\newcommand{\fig}[1]{figure~\ref{fig:#1}}

\begin{document}

\begin{abstract}
The representation of arbitrary data in a biological system is one of the most elusive elements of biological information processing. The often logarithmic nature of information in amplitude and frequency presented to biosystems prevents simple encapsulation of the information contained in the input. Criticality Analysis (CA) is a bio-inspired method of information representation within a controlled self-organised critical system that allows scale-free representation. This is based on the concept of a reservoir of dynamic behaviour in which self-similar data will create dynamic nonlinear representations. This unique projection of data preserves the similarity of data within a multidimensional neighbourhood. The input can be reduced dimensionally to a projection output that retains the features of the overall data, yet has much simpler dynamic response. The method depends only on the rate control of chaos applied to the underlying controlled models, that allows the encoding of arbitrary data, and promises optimal encoding of data given biological relevant networks of oscillators. The CA method allows for a biologically relevant encoding mechanism of arbitrary input to biosystems, creating a suitable model for information processing in varying complexity of organisms and scale-free data representation for machine learning.
\end{abstract}

\title{Criticality Analysis: Bio-inspired Nonlinear Data Representation}

\author{Tjeerd V. olde Scheper}
\affiliation{School of Engineering, Computing and Mathematics, Oxford Brookes University\\Wheatley Campus, Oxford, OX33 1HX, United Kingdom\\\url{tvolde-scheper@brookes.ac.uk}}

\maketitle

\section{Introduction}
The aim in methodologies applied to feature extraction for machine learning is to find separation boundaries within the data sets such that categorisation algorithms can readily classify the samples. Different methods rely on specific approaches, based on probabilistic or deterministic principles. In general, the most effective approaches are those that emphasise the common features shared within the class, and understate features that separates class members, and ideally perform the opposite for between-class comparisons \cite{Ribeiro2016}. Most commonly these problems form part of some regression where multimodal input is reduced to a few representative variables \cite{Bishop2006}. Here, it is proposed that these types of problems are common in biological systems, and that a bio-inspired approach based on simple connected networks of nonlinear controlled biochemical oscillators is capable of creating nonlinear representation spaces that facilitates classification greatly. The network is based on individual units from biochemical origins that are nonlinear and can exhibit chaotic behaviour. These units are then controlled using the Rate Control of Chaos (RCC) method \cite{OldeScheper2017b} which can stabilise the dynamic behaviour into steady state or periodic orbits. The method ensures that each individual unit can maintain stability even when perturbed externally. The cumulative effect of the total network shows stable dynamic behaviour that shares properties with Self-Organised Critical systems \cite{OldeScheper2022a}.

The concept of Self-Organised Criticality is the notion that a nonlinear dynamic system can be at an equilibrium or critical point such that a small perturbation may cause it to transition from one state to another state. The underlying dynamic behaviour of the elements that form such a system is based on local dynamic behaviour that contributes to global stable states. Such a mechanism has previously been conjectured to underlie biological dynamic behaviour \cite{Mora2011,Kaufman1993}. This requires, however, a control method that is capable of ensuing the dynamic stability of those units; even if they are perturbed externally. These units would, without such a control method, result in unstable dynamic behaviours, such as is the case in spatiotemporal chaos. The use of RCC allows such a network to maintain regularity even when perturbed, and from this network it has the emergent property that contains all three key features of a Self-Organised Critical system, namely non-trivial scaling due to external perturbation, spatiotemporal power-law correlation with respect to the total behaviour of the network, and self-tuning to a critical point \cite{Watkins2016}. The concept of stability and SOC despite external perturbations allows a representation of input resulting in a uniquely dynamic consistent representation which forms the basis of the Criticality Analysis method.

The resulting network of oscillators is adaptive, and dynamically stable, allowing a system that is therefore suited to act as a deterministic reservoir of oscillators, capable of receiving input in a deterministic and consistent manner that can be apparently scale-free. This reservoir of oscillators is, even in its most basic representation, capable of uniquely creating a stable and consistent representation of input, especially if this is in itself dynamic. It has been known that a random reservoir \cite{Natschlaeger2002b,Steil2004,Dutoit2008} is able to represent a large number of input states, such that for an extremely large number of input units, any given input can be represented. The advantage of the Critically Analysis method is that the network states are deterministic, in the sense that the resulting states are consistent with the presented input, and each input will result in the same state. Furthermore, similar input will result in nearby states in the dynamic sense, where the representation spaces is dynamically separated into spatial domains. The CA method will also provide the ability to change the dynamic behaviour of the network in frequency or other dynamic properties. Additionally, unlike in the case of a echo state machine or reservoir computing approach \cite{Jaeger2002b}, the number of units is fairly small, and the number of oscillators and their connectivity affects the dynamic response, but not so much the representation, allowing scale free consistent representations as will be shown in the results.

The relevance of this approach for machine learning is that such a method will allow suitable representation of both static and dynamic data for classification, although the nature and number of oscillators needed to represent any given data set is still very much under study. It is clearly the case that no arbitrary dataset is ideally suited specifically for any given network of oscillators, for which idealised classifiers can be defined, but it will allow an universal approach to the data, that is not possible without the CA representation. The CA representation permits static, dynamic and multimodal data to be directly represented alongside each other, independently from  their range or variance. As will be shown, the CA approach also allows scale-free representation, and regressed representation of complex data, on which classification may readily be performed. The approach is in itself not a classifier, although it shares similarities to some clustering methods, but permits consistent and deterministic representations on which a classifier can be applied. Existing methods for regularisation focus on the reduction of outliers, such as irregular, or low frequency occurring data values. The CA method does not seem to require this, as the resulting representation is by its nature regularised into the stable deterministic oscillation or steady state.

The relevance of the approach to biological representation should also be emphasised. The problem of any given biosystem of receiving data with high variance and logarithmic scales into a limited representation capable system of receptors or biochemical dynamic systems  should not be underestimated. It has been known that a biosystem is required to be able to deal with such problems in a deterministic and reproducible manner \cite{Kaufman1993}. The proposed mechanism of a network with Self-Organised Criticality properties with consistent scale-free representation based on relatively simple biochemical oscillators would allow even the smallest nontrivial biosystem to receive external input perturbation such that its own integrity and regularity are not affected, and it is able to respond in a consistent and scaled manner to those inputs.

\section{Methods}
The Criticality Analysis method is based on a network of Rate Control of Chaos (RCC) controlled nonlinear oscillators \cite{OldeScheper2017b}. These oscillators are, in effect, stable oscillators due to the RCC method. However, the parameter space of the oscillators is in the chaotic domain. The oscillators are still weakly chaotic, i.e. when perturbed they will return to a periodic state, but not necessarily to the same orbit as before the perturbation due to the control. Because the RCC oscillator is deterministic, it will always go to the same orbit with the same perturbations, and the perturbation that is somewhat similar will result in a nearby, but uniquely different orbit. The network can be effectively controlled by these perturbations for this reason.  It has already been shown that a network of such controlled oscillators is capable of stabilising chaotic and noisy oscillators, even when the instability is due to complexity, such as is the case in spatiotemporal chaos \cite{OldeScheper2017b}.  Furthermore, a network of these RCC controlled oscillators shows scale free behaviour, emerging as higher order relations between the perturbations and the total system response. This depends on the domain of control of the oscillators, and their connectivity, as would be expected \cite{OldeScheper2022a}. 

In this paper,  a model of a bienzymatic oscillation is used developed by Berry \cite{Berry2003}, which is described by \eqref{berry:m} to \eqref{berry:gmod}. This model has previously been shown to be controllable using the Rate Control of Chaos (RCC) method \cite{Scheper2008}. The model describes two enzymes that control the formation of extracellular matrix $m$ from soluble filaments $f$. The proteinase $p$ transforms the matrix into filaments, and conversely transglutaminase $g$ converts the filaments into the matrix.  Extracellular matrix is also produced $r_{im}$ at a constant rate, and each protein decays proportionally to $p$.  $r_{im}$ is the bifurcation parameter that causes the system to be chaotic within specific domains. By applying the Rate Control of Chaos, which is described by the quotient $q_f$ \eqref{berry:qf}, and the two control functions $\sigma_p$ \eqref{berry:sp}, and $\sigma_g$ \eqref{berry:sg}, the system remains in controlled stable orbits for a wide range of values of the bifurcation parameter.  Within the subsequent simulations, the model is usually shown as a time series of the main variables $m$ and $f$ or as phase space plots of $f$ (x-axis) versus $m$ (y-axis). Additionally, the total summed behaviour of the main variables will be referred to as $M$ and $F$ (capitalised).

\begin{align}
 \label{berry:qf} q_f&=\frac{f}{f+\mu_f}\\
 \label{berry:sp} \sigma_p(q_f)&=f_p\,e^{\left(\xi_p\,q_f\right)}\\
 \label{berry:sg} \sigma_g(q_f)&=f_g\,e^{\left(\xi_g\,q_f\right)}\\
  \label{berry:m} \frac{d\,m}{d\,t}&=k_g \frac{f\,g}{K_G+f}-\frac{m\,p}{1+m}+r_{im}\\
 \label{berry:f} \frac{d\,f}{d\,t}&=-k_g \frac{f\,g}{K_G+f}+\frac{m\,p}{1+m}-\frac{f\,p}{1+f}\\
 \label{berry:pmod} \frac{d\,p}{d\,t}&=\sigma_p(q_f) \gamma \frac{f^n}{K^n_R + f^n}-k_a\,p^2\\
\label{berry:gmod} \frac{d\,g}{d\,t}&=\sigma_g(q_f) \beta \frac{f^l}{K^l_S + f^l}-k_{deg} \frac{g\,p}{K_{deg}+g}
\end{align}

The Berry model parameters are as follows; $\gamma=0.026, \beta=0.00075, K_R=4.5,  K_S=1, K_G=0.1, K_{deg}=1.1, k_g=k_{deg}=0.05$, $k_a=\frac{k_{deg}}{K_{deg}}=0.0455$ and the Hill-numbers $l=n=4$. For different values of the bifurcation parameter $r_{im}$ in \eqref{berry:m}, the model exhibits a wide range of dynamic behaviour, including periodic cycles, bistability and chaos \cite{Berry2003}, if left uncontrolled. The RCC control may be tuned for different amounts of control, in most cases (unless otherwise stated), the control parameters are $\mu_f=2$, $f_p=1$, $f_g=1$, $\xi_p=-3$, $\xi_g=-3$. Deterministic external input $\epsilon$ is added to the $r_{im}$ parameter as in equation \eqref{sum:rim}, and it is also used to connect the different oscillators together using a scaled relative contribution from the other $n-1$ oscillators (without self-connections):

\begin{align}\label{sum:rim}
	r^i_{im}&=\sum^n_{k=1, k\neq i} w_k\, m_k + \epsilon
\end{align}

\noindent where $w_k$ the connectivity strength for the oscillator which is fixed for these simulation, and importantly, remains within the chaotic domain. $\epsilon$ is a variable that represents the external data input and matches different attributes of the data provided to the input oscillators. These perturbations are fixed for a number of time steps of the evolution of the total system to allow the system to stabilise into a new orbit due to the new data presented. It should be made clear that the system will remain in the new stable orbit until the input is changed again, when it will move to yet another orbit.

The Rate Control of Chaos method is furthermore used to provide dynamic scaling of the input variables. This can be employed to ensure that the input data is normalised and relies on the relative change of the input $\epsilon$ to perturb the input oscillator proportionally, which is indicated by $\epsilon_i$ for each input variable in the range $\{\epsilon_i,\ldots,\epsilon_n\}$ for any $n$ attributes of the data. This dynamic normalisation is described by the following equations.

\begin{align}
\label{sum:rimnorm1}	q^i_\epsilon&=\frac{\epsilon_i}{\epsilon_i+\mu^i_\epsilon}\\
\label{sum:rimnorm2}	\sigma^i_\epsilon&=f^i_\epsilon e^{\left(\theta^i_\epsilon\, q^i_\epsilon\,q_i\right)}\\
\label{sum:rimnorm3}	r^i_{im}&=\sum^n_{k=1, k\neq i} w_k\, m_k + \sigma^i_\epsilon \epsilon_i
\end{align}

where $\epsilon_i$ is the external input that represents a data point, $\mu^i_\epsilon$ is the maximal value of the data (which is predetermined and constant), $f^i_\epsilon$ is a constant scalar, and $\theta^i_\epsilon$ is the RCC control variable as described before, with mostly $f^i_\epsilon=1$ and $\theta^i_\epsilon=-3$.  $q_i$ is the proportional RCC scaling for oscillator $i$ to which the input is presented (cf. \eqref{berry:qf}), this allows the input to be scaled in relation to the oscillator and also makes static data (such as a constant value) a dynamic variable. The RCC control is finally used to scale the input variable $\epsilon_i$ itself, and added to the $r_{im}$ parameter of oscillator $i$. 

The dynamic Hebbian learning rule used for the simulations where the network is adjusted based on the cross-correlation of their activities, is described by the following equations.

\begin{align}
\label{dh:qfi} qf_i&=\frac{f_i}{f_i+\mu_i}\\
\label{dh:qfj} qf_j&=\frac{f_j}{f_j+\mu_j}\\
 \label{dh:1}  w_{ij}&=\theta_i e^{\left(\alpha_i\,qf_i\,qf_j\,\sign(qf_j-qf_i)\right)}
\end{align}

\noindent where the quotients $qf_i$ and $qf_j$ represent the proportion of filament as in equation \eqref{berry:qf} for the respective oscillators $i$ and $j$, and the weight projecting from $i$ to $j$ is adjusted according to \eqref{dh:1} which strengthens the connection if the difference in these quotients is positive (i.e. the target is stronger than the source) and reduces the strength when this is not the case. Because the value of $f_i$ and $f_j$ may change over time, the weights may adapt to the changing behaviour of the system. Typical values for the learning parameters are $\alpha_{i}=-1$ and $\theta_i=1$.

The second model used is a four dimensional extension of the classical 3-dimensional Lorenz model that incorporates an additional feedback loop by Wu et al. \cite{Wu2009}. The model can exhibit chaos and hyperchaos, and is described by the equations \eqref{wu:x}-\eqref{wu:w}:

\begin{align}
 \label{wu:qx} q_x&=\frac{x}{x+\mu_x}\\
 \label{wu:qy} q_y&=\frac{y}{y+\mu_y}\\
 \label{wu:qz} q_z&=\frac{z}{z+\mu_z}\\
 \label{wu:sx} \sigma_x&=e^{\left(\xi_x\,q_x\,q_z\right)}\\
 \label{wu:sy} \sigma_y&=e^{\left(\xi_y\,q_x\,q_z\right)}\\
 \label{wu:sz} \sigma_z&=e^{\left(\xi_z\,q_x\,q_y\right)}\\
 \label{wu:sw} \sigma_w&=e^{\left(\xi_w\,q_y\,q_z+\omega \right)}\\
  \label{wu:x} \frac{d\,x}{d\,t}&=a (y-x) + e \sigma_x y z + k w +D_x\\
 \label{wu:y} \frac{d\,y}{d\,t}&=(c x - d y - \sigma_y x z) + D_y\\
 \label{wu:z} \frac{d\,z}{d\,t}&=(\sigma_z x y - b z) +D_z\\
\label{wu:w} \frac{d\,w}{d\,t}&=(\rho y + f \sigma_w y z) +D_w
\end{align}

\noindent where $a=56$, $b=16$, $c=49$, $d=9$, $e=30$, $f=40$, $k=8$, $r=-600$ for the system parameters, and $\mu_x=1000$, $\mu_y=100$, $\mu_z=100$, $\mu_w=3000$, $\xi_x=-1.5$, $\xi_y=-1$, $\xi_z=-1$, $\xi_w=-1.5$, $\omega=0.055$ for the RCC parameters in the RCC equations \eqref{wu:qx} - \eqref{wu:sw}. The external input to the oscillators is provided by $\rho=-100 \epsilon$ with $\epsilon$ as the perturbation value representing data for each oscillator. Additionally, the $n$ oscillators are connected to each other using the diffusion terms $D_x$, $D_y$, $D_z$, $D_w$ where each term is simply the weighted sum of the corresponding variables for the oscillators, e.g. $D_x=\sum^n_{k=1} w_k x_k$, with optional weighting term $w_k$. Note that the bifurcation parameter $r$ is chosen in the hyperchaotic domain, see figure 1 in Wu et al. \cite{Wu2009}.

The models were simulated using the EuNeurone numerical integration software, which is available on Zenodo \cite{OldeScheper2021a}. The results can nevertheless be readily simulated using other fixed-step numerical integration tools. The numerical integrators used were the standard fixed step integration Runge-Kutta RK4,  and Fehlberg-RK algorithms \cite[pages 363-366]{GSL2009}. Results were then exported into Hierarchical Data Format  5 (HDF5), subsequently analysed, and plotted using Matlab. For each figure in the manuscript, a corresponding model file is available from Zenodo \cite{OldeScheper2022b} that allow reconstruction of the results.

The datasets used in these experiments are the standard, unmodified versions of the classic Iris data set (three classes, 4 attributes, 120 samples, numerical),  the Wine dataset (three classes, 13 attributes, 178 samples, multivariate), and the Bonemarrow dataset (unclassified, 36 attributes, 172 samples) from the UCI Machine Learning Repository \cite{Dua2017}. The order of the samples were randomised before input to the oscillators, and the category labels were not presented to the networks, these were only used to determine the class after the simulations. With the Bonemarrow dataset, all incomplete samples were excluded, and three of the attributes were unused. The main aim of this dataset is to test for possible relations between some attributes and survival \cite{Sikora2019}.
It is important to note that the data in these models is used to generate the nonlinear representation spaces, not primarily to provide categorisation directly. These models are shown to allow a biologically relevant mechanism to represent dynamically changing, as well as static, data to be represented in a manner that facilitates categorisation and is insensitive to scale, but sensitive to variability within the sample data set.

\section{Results}
\subsection{Simply Perturbed Berry Model}
\begin{figure*}[ht]
\centering\includegraphics{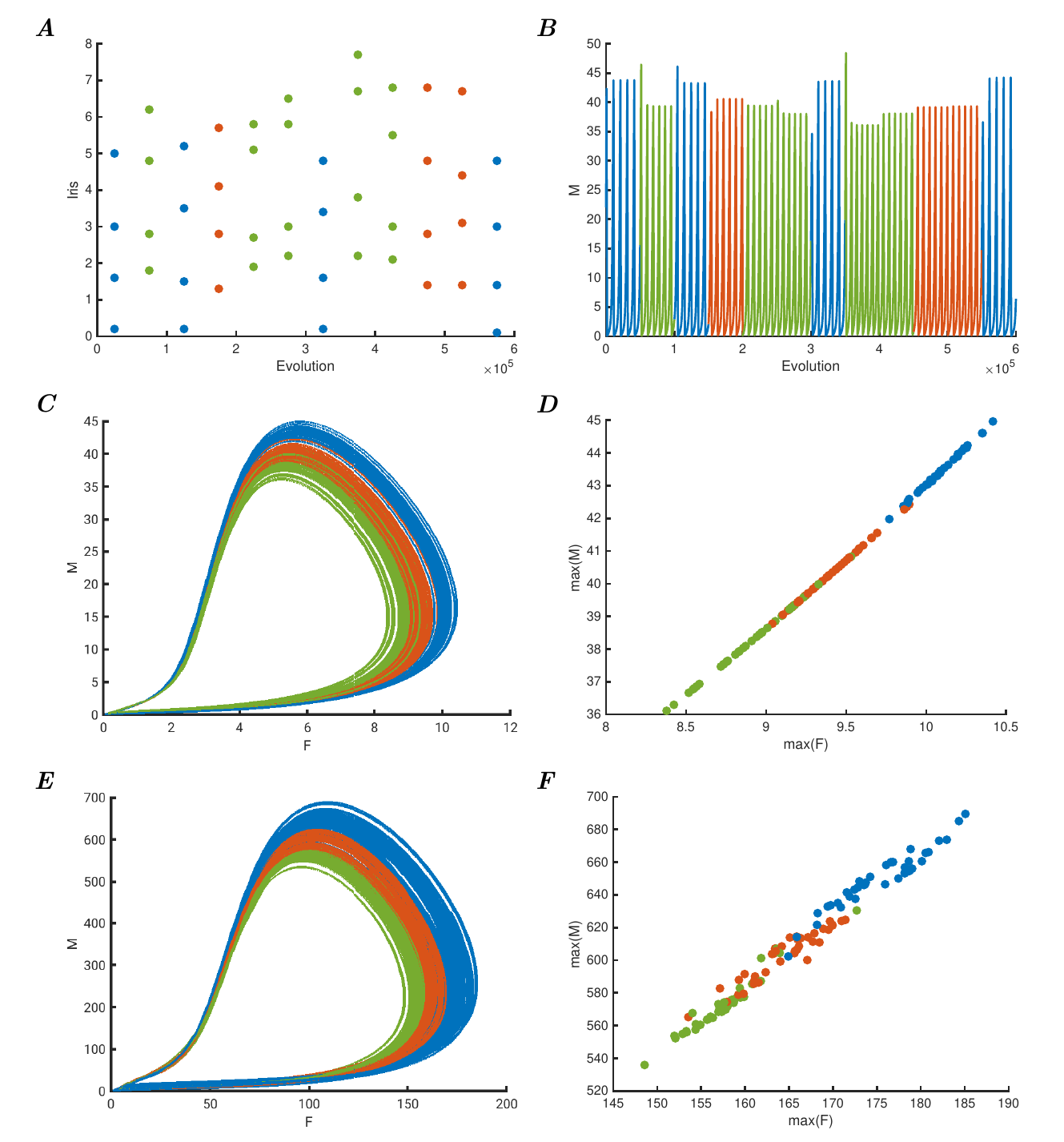}
\caption{\label{fig:1}
\apanel{A} First 12 samples of the Iris data set as presented to the 4 oscillator network. \apanel{B} Total $M$ resulting from the perturbations caused by the Iris dataset of the network. \apanel{C} Total $F$ versus $M$ of the 4 oscillator Iris perturbed network showing different dynamic domains representing categories. \apanel{D} Maxima of $F$ versus $M$ of the figure in panel \apanel{C} demonstrating possible linear separation bounderies. \apanel{E} Iris perturbed phase space plot of $F$ versus $M$ of the 64 oscillator network, showing size independent constant representation of the data. \apanel{F} Phase space plot of the maxima of the 64 oscillator Iris perturbed network, showing consistent representation of the data.
}
\end{figure*}

Let us first consider the simplest representation case, where the four dimensional Iris dataset is projected onto a small network of four Berry oscillators. These oscillators are RCC controlled, as previously described, and have mutual weights of $w_k=0.0005$, and the iris data is scaled by $0.00025$ when added to equation \eqref{sum:rim} as the parameter $\epsilon$. In \fig{1}\apanel{A} is shown the first twelve data points of the Iris data sets as the four values presented to the network. Each data point is clamped to the input for 50,000 evolution time steps where the step size is $0.1$. In the next panel \fig{1}\apanel{B} is shown the dynamic response of the network of those twelve data points, as the total amount of $M$ of the four oscillators, colour coded to the correct class. The class type itself is not used by the network as such. In panel \apanel{C} is shown the resulting dynamic response due to the data perturbation of the entire iris data set as a phase space representation of the total $F$ versus $M$. Here the blue class (\textit{Iris setosa}) causes the largest perturbations, followed by the red class (\textit{Iris virginica})  and finally the green class (\textit{Iris versicolour}). Note that the last two are the classes that require nonlinear separation boundaries to classify \cite{Dua2017}. To allow for readily classification, although the aim here is create dynamic representation spaces not optimise for classification, the maxima of each of the total oscillations is plotted in panel \apanel{D}. It should be clear that the representation is not perfectly separable given the data, but that even a simple perceptron classifier can be readily trained to perform reasonably well on this representation using two simple linear boundaries. This representation of the Iris data is due to the dynamic response of the network to the perturbations that represent the data points, and is emphasised by the criticality properties of the controlled chaotic system. Additionally, this dynamic response requires only relevant input, and no training for the representation is necessary.

To demonstrate this interesting property of criticality in response to deterministic perturbations, the network is extended to 64 RCC controlled Berry oscillators. The Iris data is presented to the network as input to only the first four oscillators with scaling of $0.005$, and the network weights are reduced to $w_k=0.00005$. Apart from the reduction of overall connectivity strength and the number of oscillators, the model remains the same. Due to the critical response, i.e. the ability to change state from one orbit to another due to small perturbations (in this case the data), the network response is qualitatively similar to the small network of four oscillators, as can be seen in \fig{1}\apanel{E}, where the phase space plot of the orbits colour coded to the categories is shown. Subsequently, in \fig{1}\apanel{F} is shown the maxima of these orbits on which a classifier can readily be trained.

\subsection{Oscillators with Dynamic Hebbian Learning}
\begin{figure*}[ht]
\centering\includegraphics{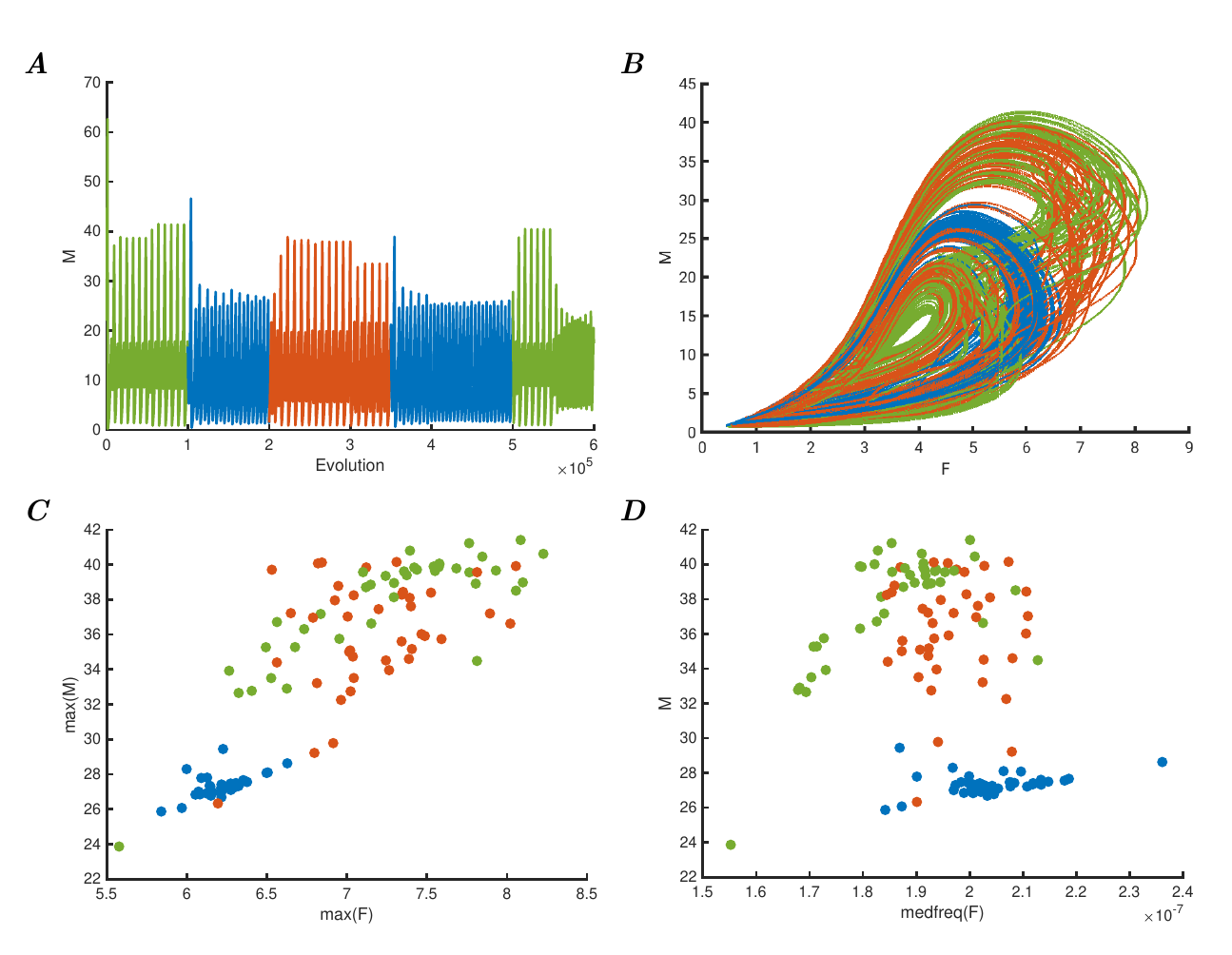}
\caption{\label{fig:2}
\apanel{A} Evolution in time of 12 samples from the Iris data set input to a network of oscillators with dynamic Hebbian learning. \apanel{B} Phase space plot of the total $F$ versus $M$ of all the Iris samples, showing changes in dynamic response due to the perturbations causing adaptive learning. \apanel{C} Maxima of total $F$ versus $M$ of all the Iris samples in the network, showing poor separation on amplitude. \apanel{D} Median frequency of the total $F$ versus the maximal $M$ showing that separation is possible due to the changing dynamics of the orbits, as is shown in panel \apanel{A}.}
\end{figure*}

It is well understood that the connectivity weights in a network of oscillators can affect the behaviour of the system, and to explore this property a subsequent set of experiments were devised. A network of four oscillators is adapted with the dynamic Hebbian learning rule according to equation \eqref{dh:1}. The weights are responding to the changing behaviour of each oscillator from $i$ to $j$ where the latter is the target and the former the source. The external data input is provided as in the previous models, but with enhanced input strength of $0.005$ as described before. The learning parameters are $\alpha_{i}=-1$ and $\theta_i=1$ for all connections in this experiment.
 The results are shown in \fig{2}\apanel{A}, where twelve orbits of the total $M$ are shown in time. In the subsequent panel \apanel{B} is shown the phase space plot of $F$ versus $M$ of the entire Iris data set. In panel \apanel{C} is shown again the maxima of these two variables. At first glance, it may appear that the representation in this manner makes it harder to classify, however, as can be seen from the first panel in the \fig{2}\apanel{A}, the shape of the oscillation has changed rather than the amplitudes. This shows that the connectivity of the critical network of oscillators affects the dynamic behaviour of the system, rather than the individual amplitudes or peaks of each of the periods. To classify such dynamic behaviour, in a meaningful biologically relevant sense, would require the subsequent classifier to be able to receive the dynamic input of the representation and separate them as different evolutions in time. To illustrate this further, the Median Frequency of each total oscillations for the total value of $F$ was estimated, and plotted against the maxima of the total value of $M$. This is shown in the panel \apanel{D}, where the adjusted mean frequency estimate can be used in combination of the maxima to separate the classes, in a non-optimal but suitable manner. The importance is not the efficacy of the classifier, but the ability to represent complex data in different ways that are meaningful for continuous dynamic systems such as biosystems to permit classification in the first place.

\subsection{Deterministic Reservoir Computing}
The property of changing dynamics based on external perturbation may be exploited in a different manner, as part of a representative state based on these perturbations. Such a reservoir of dynamic behaviour requires then a readout layer for categorisation. To explore this aspect, another network was devised that contained six Berry type of oscillators with two additional external oscillators that may act as readout units. These two external oscillators are configured similarly to the six reservoir units. In this simulation, the network of six oscillators that form the reservoir do not contain any learning, similar to the four units in the first model, but the learning rule is only applied to the connections of all six units to the readout units \cite{Steil2004}. The data is provided to four of the six reservoir units by equation \eqref{sum:rim}. The input to the readout units seven and eight are given by

\begin{align}
	r_{im}^7&=\phi_7 \sum^{n=6}_{k=1} w_k\, F + m_8\\
	r_{im}^8&=\phi_8 \sum^{n=6}_{k=1} w_k\, M + m_7
\end{align}
where $F$ and $M$ are the summed values of the six $f$ and $m$ variables respectively, and $\phi_7=\phi_8=0.00015$ as scalars. $w_k$ represent the weight from each of the six units to unit $7$ and $8$ in each equation and is determined by the dynamic Hebbian learning rule \eqref{dh:1}. In other words, the input to each unit here is the summed weight in proportion to the total value of $F$ of the reservoir for unit $7$, and in proportion to the total value of $M$ for $8$. In addition, the value of $m_8$ is added to unit $7$ and vice-versa to reflect the effect the units have on each other. The parameter values are the same as in the previous models, except that the external input is scaled to $0.00225$, and the fixed connections in the reservoir are $0.0002$. The learning parameters are $\alpha_{i}=-3$ and $\theta_i=3$ for the two readout units. 

\begin{figure*}[ht]
\centering\includegraphics{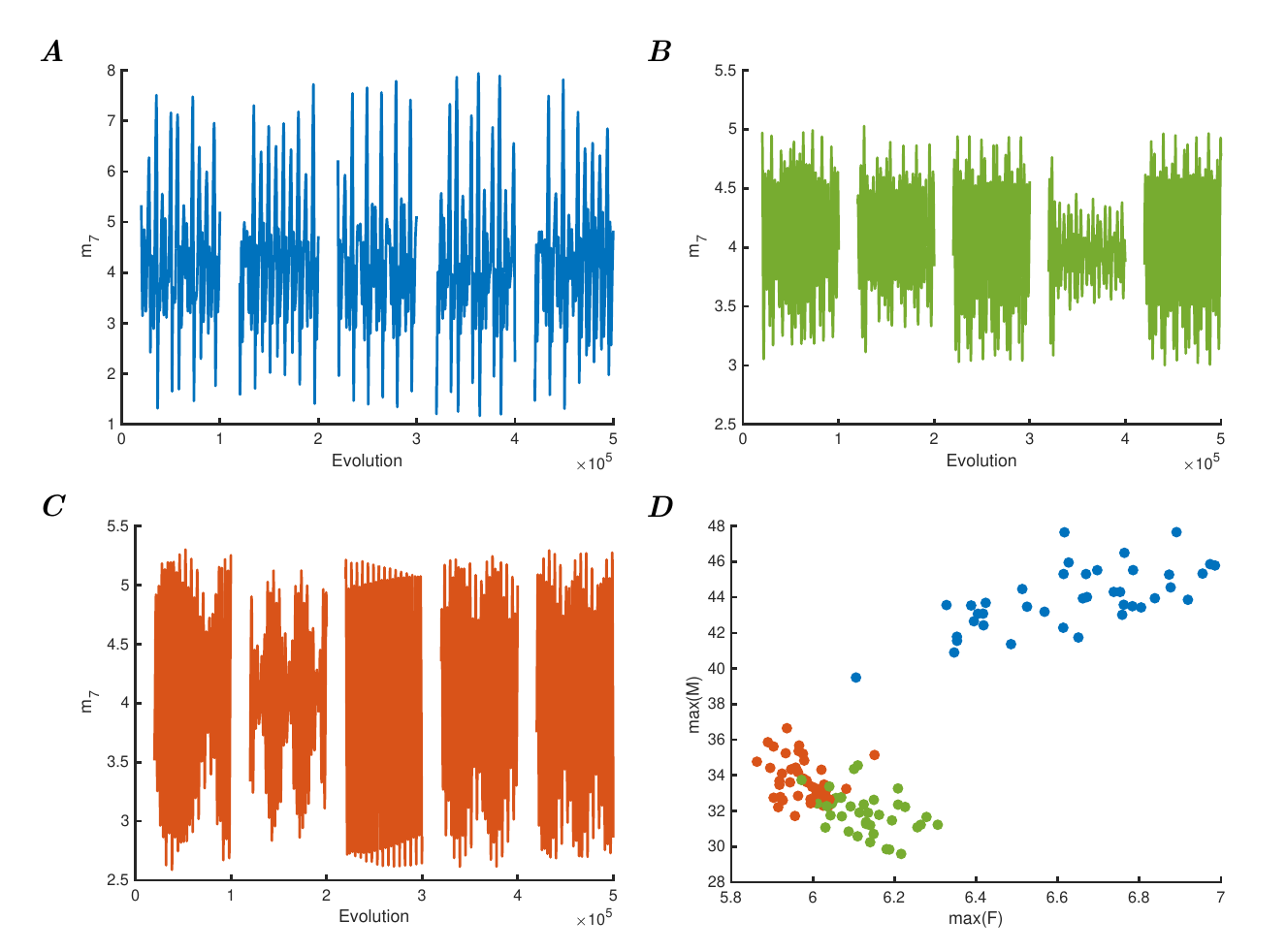}
\caption{\label{fig:3}
\apanel{A} Orbits of 5 samples of the first class in the Iris data set as represented by the read-out unit 7 in time. \apanel{B} Orbits of 5 samples of the second class in the Iris data set as represented by unit 7 in time. \apanel{C} Same as previous panels, but for the third class. \apanel{D} Phase space plot of the maxima of total $F$ and $M$ of the six reservoir units (i.e. excluding the read-out units, 7 and 8).}
\end{figure*}

\begin{figure*}[ht]
\centering\includegraphics{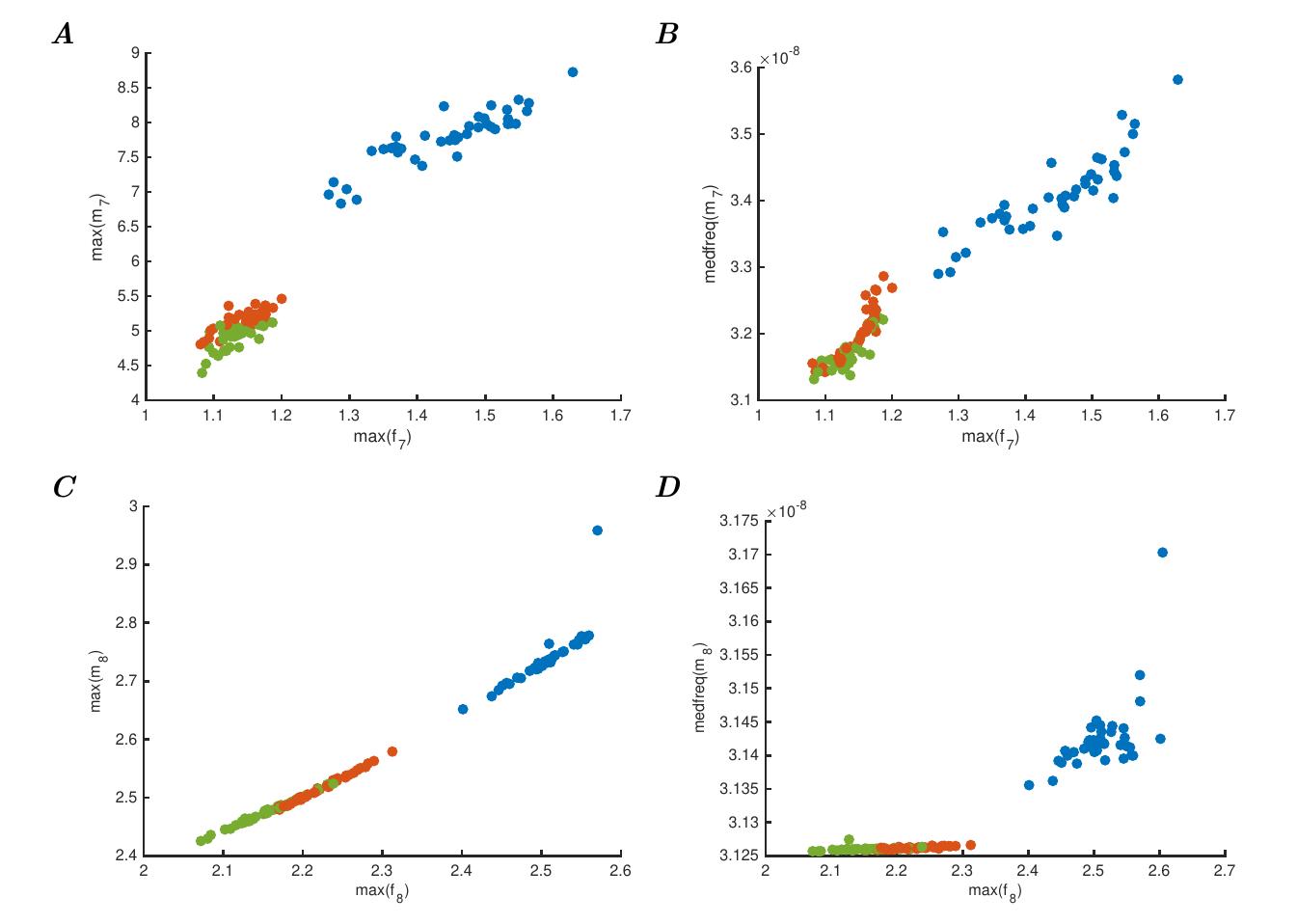}
\caption{\label{fig:4}
\apanel{A} Phase space plot of the maximal $f$ and $m$ from the read out unit 7. \apanel{B} Phase space plot of the maximal $f$ versus the median frequency of $m$ of unit 7. \apanel{C} Phase space plot of the maximal $f$ and $m$ of the read out unit 8. \apanel{D} Phase space plot of unit 8, showing maximal $f$ versus the median frequency of $m$.}
\end{figure*}

The result of presenting this reservoir with the Iris data set is shown in \fig{3}. In panels \apanel{A} to \apanel{C} are shown five arbitrarily chosen dynamic responses to each of the three categories as observed by the readout unit $7$ in time. Because the connectivity strength is higher than in the first experiment, the network responds with different oscillations due to the perturbations of each data point. As can be seen, there are significant differences in these responses, and although some points are not easily defined, they are each recognisably representative for their class. In panel \apanel{D} is shown the maxima of all these oscillations for the Iris data set, which can be compared to the single orbit oscillations that characterised the first model (described in \fig{1}). Furthermore, by comparing the individual readout units $7$ and $8$ in \fig{4}, it is clear that these units independently show different aspects of the reservoir behaviour. The panel \apanel{A} shows the maxima of unit $7$, and \apanel{C} the maxima of unit $8$. The first shows the same characteristic behaviour as the overall reservoir (compared to \fig{3}\apanel{D}), and the panel \apanel{D} shows the characteristic behaviour as comparable to the original model in \fig{1}\apanel{D}. It can therefore be concluded that the readout units show a dynamically reduced but characteristic behaviour of the entire reservoir, i.e. each individual oscillator shows the same behaviour for which the entire network is required to produce. The exact nature of the properties of the classifier required to actually separate these classes using either of the readout units (or both), is not essential for this argument, but it is clearly much easier to classify on one oscillator's behaviour rather than a network, or indeed the data itself. To complete the comparison of this reservoir computing approach, the Median Frequencies of the two readout oscillators is also determined and is shown in \fig{4}\apanel{B} for unit $7$ and in \apanel{D} for unit $8$, showing that the median frequency versus the maxima of $f_7$ and $f_8$ respectively. These are different again from the previous model experiment (in \fig{2}) due to the lack of adaptation in connections and the overall readout of the network.

\subsection{Simply Perturbed Wu model}
\begin{figure*}[ht]
\centering\includegraphics{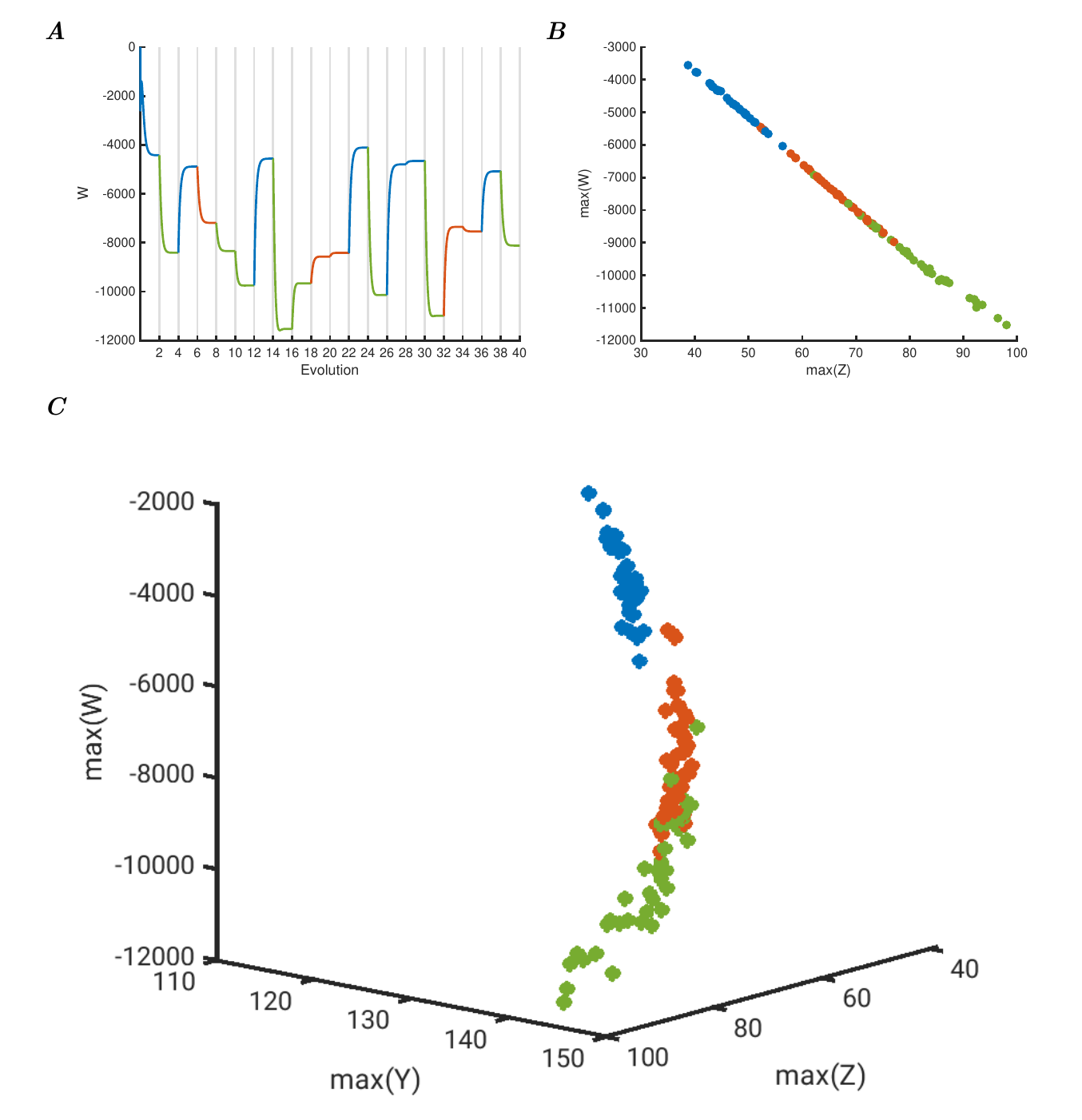}
\caption{\label{fig:5}
\apanel{A} Network of four Wu oscillators with Iris data input, showing the evolution of 20 samples with total $W$ of the network. The network does not oscillated and converges to steady states, as shown in the plot where the transients end at a specific value of total $W$. \apanel{B} Phase space representation of the maximal total $Z$ versus $W$ showing the steady state representations in two dimensions allowing separation. \apanel{C} Three dimensional representation of the maxima of total $Y$, $Z$ and $W$, with the Iris data presented to the four Wu oscillator network, showing different representation in multiple dimensions. }
\end{figure*}

The previous experiments may give the impression that the choice of the Berry model is essential to these results, but apart from the requirement of the RCC control to provide the critical state, different models can be used. As illustration, the Iris data was presented to a network of Wu models. These oscillators are an extended version of the classical Lorenz model. The parameters are described in the method  section, and the results can be seen in \fig{5}. In panel \apanel{A} is shown $W$ (the summed value of the four $w$ variables of the four models) in time for 20 samples, colour coded for each category. Notice that the system is now not oscillating, but moves to representative controlled steady states. In the next panel \apanel{B} is shown the phase space plot of the maxima of $Z$ versus $W$ (after transients have been removed), in effect showing the steady state points. The relative location of these points is comparable to the ones produced by the Berry model, due to the fact that the perturbations cause somewhat similar dynamic responses, even if the quantitative response is very different. That this is the case for a specific perturbation and made visible for a specific projection of the representation space can be seen in panel \apanel{C}, where the phase space plot of the maxima of three of the four system variables is shown, with the relative representation clearly visible. The simple perturbed model approach is appropriate for the Iris data set, and would allow readily categorisation based on the network response. 

\subsection{Dynamic Normalisation}
\begin{figure*}[ht]
\centering\includegraphics{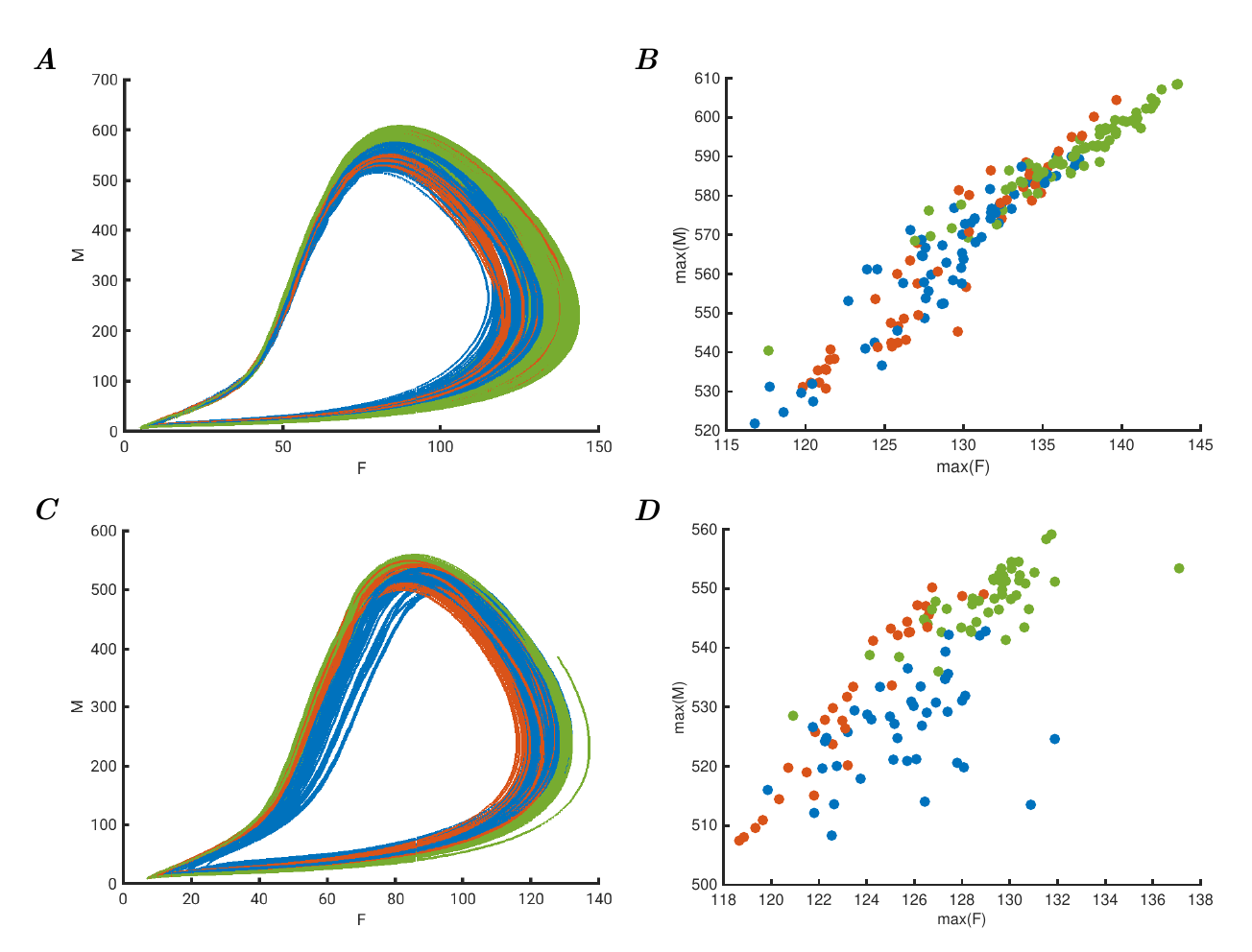}
\caption{\label{fig:6}
\apanel{A} Network of 64 Berry oscillators with unmodified, scaled input from the 13 attribute Wine data showing phase space of all samples as total $F$ versus $M$. \apanel{B} Maximal $F$ versus $M$ of all the Wine samples, showing poor inherent separation based on amplitude. \apanel{C} Same 64 oscillator network, but with dynamic normalisation function to adjust the input representation to the network, showing the phase space of total $F$ versus $M$. Note that the position of the samples in the nonlinear representation has changed. \apanel{D} Maximal $F$ versus $M$ of the dynamic normalised Wine data of all samples, showing that the normalised input has improved the dynamic nonlinear representation such that some separation will be possible.}
\end{figure*}

Normalisation is often used to ensure that the data with large variations is within the boundaries of the representation space. It allows the data to be scaled and represented as uniform as possible for the training of a classifier. Here it is shown that the use of a dynamic normalisation method has the same effect as it would have for a normal n-dimensional representation space. The standard Wine dataset with 3 classes and 13 attributes is used to show that normalisation, using the method described by equations \eqref{sum:rimnorm1}-\eqref{sum:rimnorm3}, improves the ability of the network to create a nonlinear representation space for subsequent classification. This dataset has large variations between attributes, and this can skew the ability of a classifier, which is why it tends to be normalised before processing anyway. In figure  \ref{fig:6} is shown the input of wine data into the 64 Berry oscillator network. In panel \apanel{A} and \apanel{B} are shown the data as unscaled input, i.e. as is, with scalars of $0.001$ for the first 12 attributes, and $0.0001$ for the last attribute that is about two orders of magnitude larger than the other attributes. Additionally, the network connectivity is initially randomised around a mean of $0.000035$ with a variance of $0.00001$, it is subsequently kept at the same value. This shows that the network connectivity changes the behaviour but not the dynamic response to individual data attributes. The figures show that the system behaves similarly as the Iris data (that has only 4 attributes), but is not readily classifiable on the amplitudes. Using the dynamic normalisation on top of the same approach (as described in the Methods section), but with everything else kept the same (including initial random connectivity strengths), the result shows that now the network can represent these data points as nonlinear orbits (panel \apanel{C}), whose maxima could conceivable be classified (panel \apanel{D}). 

\subsection{Representational Persistence}
After addressing the nonlinear representation available using Criticality Analysis for classification tasks, the manner in which CA behaves with respect to unknown relations within the data needs to be considered. To this end, a simulation was performed using a non-categorical data set to demonstrate that the resulting dynamic behaviour is still consistent with the input and the representation of known shared feature relations persists in the CA representation.

\begin{figure*}[ht]
\centering\includegraphics{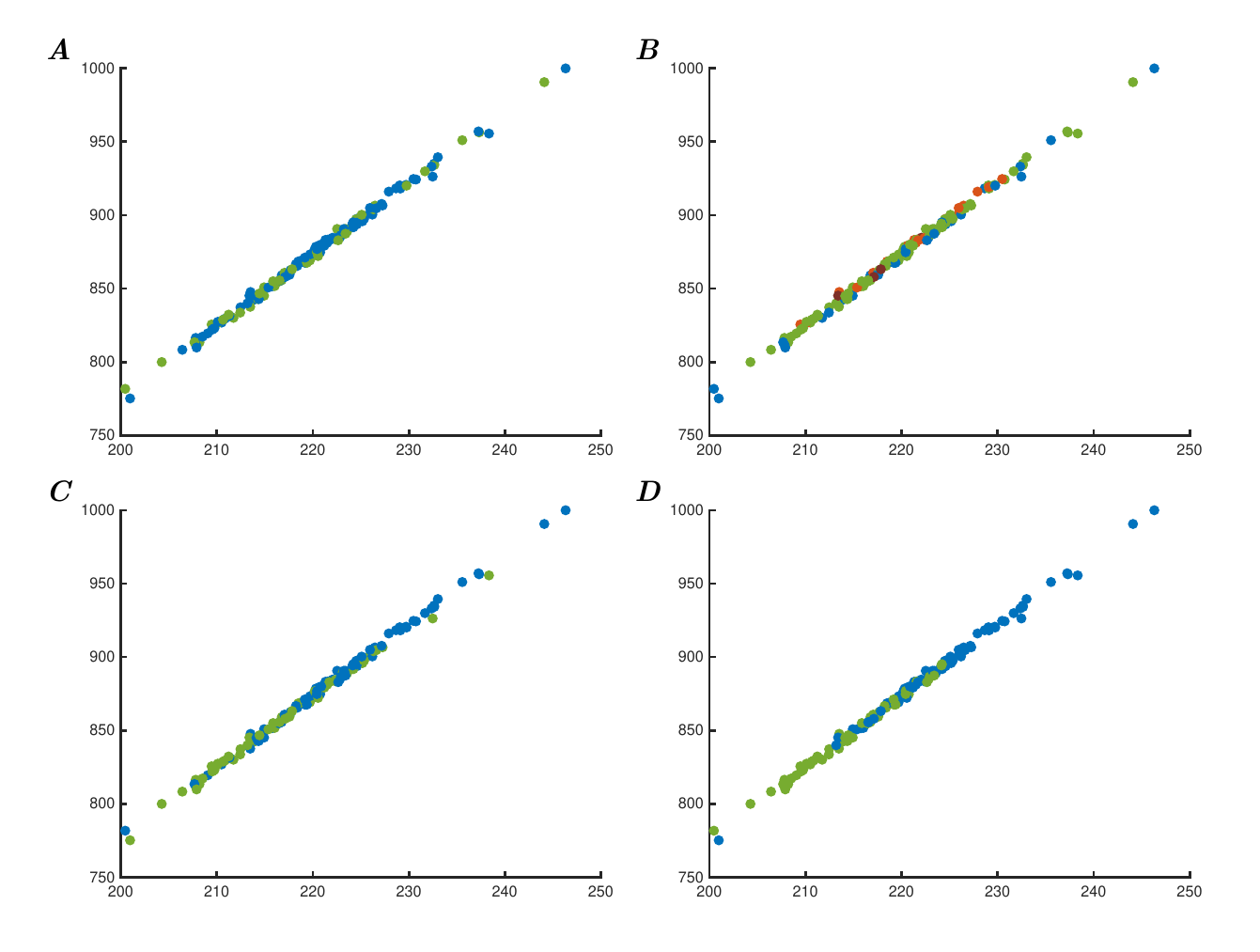}
\caption{\label{fig:7}
\apanel{A} A network of normalised input of the Bonemarrow data colour coded to survival of 64 Berry oscillators showing little relation between the two categories. \apanel{B} Same network coded to the aimed relation to the relative increase in CD34+ cells/kg showing also little relations between individuals. \apanel{C} Same network coded to the known relation with the bone marrow donor age (blue $<35$, green $\geq 35$ years old). \apanel{D} Network with colour coding to the also known relation with bone marrow recipients age (blue $<10$, green $\geq 10$ years old). The last two figures show clear grouping expressing the persistence of the age relation within the CA nonlinear representation. }
\end{figure*}

In \fig{7} are shown the nonlinear Criticality Analysis representation of 142 individuals from the Bonemarrow data set \cite{Sikora2019}. This was made using the 64 Berry oscillator network with normalisation of the 34 input features using scalars of $0.0001$ and network connectivity strength of $0.00005$. The dataset is not specifically designed for classification or regression, but to determine if some of the features increase the likelihood of survival or quality of health. Specifically, if the increased dosage of CD34+ cells per kg may extend survival. The survival status is shown in panel \apanel{A}, where no clear pattern emerges, which is consistent with the original results. The representation should show if any of the features that are similar to shared features within the dataset. If the feature is persistently responsible for a nonlinear representation state, this would show as a grouping within the nonlinear representation. In the absence of such recognisable groups, or the inability to define such groups with possible separation boundaries, the data may not contain the required information to allow this, or there is too much variability (possible due to noise). In this case, comparing the features with the representation shows that none of the features show any relation to the the categories that these represent, apart from the two age related features. It is known that both the age of the donor and the age of the recipient is a factor that exist persistently within the data representation \cite{Sikora2019}. Therefore, the resulting Criticality Analysis representation should show these relations. In panel \apanel{C} is shown the representation of the donor age showing clustering between the two classes (donor younger than 35 years, and older). Additionally, in panel \apanel{D} is shown the two age classes associated with recipients of the two age groups (recipient younger than 10 years, and older). The age relations are categorisable and demonstrate that their relations persist within the CA representation even though the other features do not.

\section{Discussion}

One of the issues with categorisation problems is that the representation of data should be in a manner such that the categorisation method requires minimal effort to train. Commonly, a large amount of effort is required to optimise this process. For many datasets this is not readily possible, due to the temporal and dynamic nature of the data, and the static nature of the learning process. Although highly optimised algorithms can be devised for specific data sets, and some significant improvement has been made to allow forms of adaptive learning, these methods are not generally applicable to dynamic data.

Biosystems need to solve problems at every level of existence, from local low-level biochemical behaviour to high level human thought. The overall idea of nonlinear critical data representation allows different levels to function reliably. The method of Criticality Analysis aims to provide a much reduced mechanism for data representation that is robust and biological relevant, as it is based on simple oscillators that can exhibit consistent and deterministic representation of input. The main requirement is the applicability of Rate Control of Chaos on a dynamic nonlinear system such that it can exhibit criticality. Here, criticality has been defined as the property of a dynamic system to change state due to small perturbations. When these perturbations are deterministic, as they represent data points, the result should therefore be deterministic behaviour as well, as can be seen in the results presented previously. 

It should be clear that the network of dynamic oscillators does not require training to create a representation. There is some need for tuning, where a minimum level of connectivity is required to ensure that the total behaviour is critical, and there is a maximal connectivity level that would destabilise the network's behaviour. This level depends on the total input of multiple oscillators, and the scalar used for the input data which were determined experimentally. However, as can be seen in figures 2, 3, and 4, training may allow the creation of representations on which a categorisation algorithm can more readily be trained. This would permit an arbitrary dataset to be presented to the network, and then a given learning algorithm may be able to cause the best representation to achieve categorisation. This does not necessarily lead to optimal representations or best categorisation, but as can be seen from the examples shown of the networks based on the Berry and Wu models, a straightforward representation can produce acceptable results with little effort.

Furthermore, the training of the reservoir can also allow the reduced representation of a more complex dataset. The algorithm used for training could readily be based on backpropagation or a similar standard neural network approach. It can therefore be argued that Criticality Analysis, if based on a adaptive suitable set of oscillators, would allow the learning of complex data to become possible even for a relatively simple organism, just by adjusting the weighting of the input of biochemical oscillators. 

The data provided to the network needs to be scaled to ensure that the total input is not destabilising the network, in a similar manner as the connectivity. The input data acts as the perturbation to the other oscillators, which explains the deterministic nature of the perturbation, as each acts as constant input to the RCC that is stabilising the critical system within each individual controlled oscillator. If the input is too strong or too weak, the network will simply not be in a critical state. The use of the normalisation approach may help here, but possibly at the price of a loss of information that may exist in the relative differences between attributes. 

Features that are persistent within the dataset are preserved within the Criticality Analysis representation. The existence of hidden or convoluted correlations are therefore not affected using the method, although it does not aid in identifying those especially.

To demonstrate the utility of the CA method for dynamic biological data, we have recently shown in a proof of concept experiment using IMU sensors that the method can be used to recognise impaired gait from the raw gait data alone\cite{Eltanani.2022}. Subsequently, we extend this proof to a large gait data set of healthy normal volunteers to show correct high performance classification using CA with SVM classifiers. Additionally, we have measured gait in children has been used in combination with CA to assess improvements in health due to clinical treatment. The papers describing these analyses are in preparation.

Lastly, the method described in this paper may be considered to be some form of a clustering algorithm. However, that would neglect the uniqueness of the representation, and the deterministic nature of the method. The relation between the data properties and the critical representation is still under investigation. It is, however, clear that not any given dataset can be best represented by any given critical model, in which case, only one model would fit all datasets. Determining which is the best critical model for a specific data representation, as well as the development of further models that have suitable properties for this type of analysis is also still under way. 

\section{Conclusion}
The results show that the nonlinear representation using Criticality Analysis does not by itself represent underlying correlations, but allows arbitrary data to be categorised using an appropriate set of oscillators with a suitable categorisation method. It should be clear that there is no fundamental reason that these oscillators are specifically useful to represent the attached data, nor are they especially suited for these datasets. The aim is to show that it is firstly possible to represent data in a dynamic and biologically relevant manner, which can also be used as a reservoir computing approach, that allows dimensional reduction without loss of representational behaviour. Secondly, that the data can be categorised without any learning in the best cases in which a simple categoriser (such as a perceptron) can readily be trained. Thirdly, that dynamically changing data is the norm within biological systems, and that the proposed mechanism may be one of the ways in which biological data may be represented in a size independent but consistent manner due to the critical nature of the network of oscillators. The Criticality Analysis method can be used to create suitable representation of complex, dynamic data for further optimised categorisation approaches, but may also shine a light on one of the most fundamental problems in biosystems, that of reliable, deterministic, and scale free data representation.\\

\noindent The author wishes to declare no conflict of interest, and that the research was performed without funding.

\bibliographystyle{elsarticle-num}
\bibliography{PreprintCriticalityAnalysisV9.bib}

\newpage
\section{Figures}

\newpage

\newpage

\end{document}